\documentclass[preprint, 10pt,
 amsmath,amssymb,  aps, prd
]{revtex4-2}
\usepackage[utf8]{inputenc}
\usepackage{hyperref}
\usepackage{slashed}
\usepackage{color}
\usepackage{graphicx}
\graphicspath{{pics/}}
\allowdisplaybreaks[1]

\begin{document}
\title{Soft gluon fields and anomalous magnetic moment of muon}
\author{Sergei Nedelko}
\email{nedelko@theor.jinr.ru}
\author{Aleksei Nikolskii}
\email{alexn@theor.jinr.ru}
\author{Vladimir Voronin}
\email{voronin@theor.jinr.ru}
\affiliation{Bogoliubov Laboratory of Theoretical Physics, JINR, 141980 Dubna, Russia}
\begin{abstract}
An impact of nonperturbatively treated soft gluon modes on the value of 
 anomalous magnetic moment of muon $a_\mu$ is studied within the mean-field approach to QCD vacuum and hadronization. It is shown that  radial excitations of vector mesons strongly enhance contribution of hadronic vacuum polarization to  $a_\mu$, doubling the contribution of one-meson processes compared to the result for ground state mesons. The mean field also strongly influences the hadronic light-by-light scattering contribution due to the   Wilson line in quark propagators. 
\end{abstract}
\maketitle
\section{Introduction}
The measurement of the anomalous magnetic moment of muon $a_\mu=(g_\mu-2)/2$ conducted by Muon $g-2$ Collaboration at FNAL~\cite{Muong-2:2021vma}
\begin{equation}
\label{a_mu_FNAL}
a_\mu^\text{FNAL}=116 592 040(54) \times 10^{-11}
\end{equation}
retained the earlier discrepancy with the Standard Model prediction (see reviews~\cite{Aoyama:2020ynm,Jegerlehner:2009ry})
\begin{equation}\label{a_mu_consensus_value}
a_\mu^\text{SM}=116 591 810(43) \times 10^{-11}.
\end{equation}
Significant fraction of the uncertainty in the SM value comes from the part described by QCD and hadrons. Since robust analytical first-principle QCD evaluations at the whole range of energies are not feasible at the moment,
hadronic contributions to the ``consensus value''~\eqref{a_mu_consensus_value} are essentially  extracted from experimental data via dispersion relations supplemented by perturbative QCD where large momentum transfer is involved. In order to gain a clearer insight into relevant low-energy dynamics, hadronic contributions to $a_\mu$ were also extensively studied with the methods of lattice QCD, Dyson-Schwinger equations, within the models of QCD vacuum and hadronization, and in the models of physics beyond the Standard Model.

In this paper, we address the specific role which soft gluon fields may play in the hadronic vacuum polarization (HVP) and light-by-light scattering (HLbL) contributions to $a_\mu$ using the mean-field approach to physical QCD vacuum.  Specifically, we will touch this issue by calculating the lowest order in the number of intermediate mesons  contributions (quark loop and one intermediate meson) within the domain model of QCD vacuum and hadronization developed in Refs.~\cite{Efimov:1995uz,Burdanov:1996uw,Kalloniatis:2003sa,Nedelko:2016gdk,Nedelko:2016vpj,Nedelko:2014sla,Nedelko:2020bba}. The driving feature of the model is the vacuum mean field represented by domain-structured configurations of almost everywhere homogeneous Abelian (anti-)self-dual gluon field which is treated nonperturbatively. These field configurations are characterized by various vacuum expectation values and can be interpreted as ``soft vacuum fields'' that reside in condensates of local operators in the context of operator product expansion. Such a mean-field approach has simultaneously provided  for static and dynamic confinement -- area law for Wilson loop and absence of  poles in the propagators of color charged fields in complex momentum plane, respectively, chiral symmetry breaking and resolution of $U_A(1)$ problem.  Upon hadronization,  the spectrum of light, heavy-light mesons and heavy quarkonia, including excited states, decay constants and form factors calculated  with a minimal set of parameters systematically agree very well with experimental data. The detailed description of various features of the approach can be found in above-mentioned articles. The mean field does not change short-distance behavior of quark, gluon and ghost propagators, but  otherwise strongly modifies propagators,  in a way consistent with the results of functional renormalization group
and Lattice QCD~\cite{Nedelko:2016gdk}.

For the present considerations, it is important that 
effective meson action describes  explicitly the  mesons with all possible quantum numbers and their interactions. In particular, it contains not only  ground state mesons but also their excited states. The absence of  poles in the momentum representation of the quark and gluon propagators in the finite complex momentum plane, i.e. confinement of dynamical color, leads to the Regge spectrum of light meson masses.  
This allows one to compute contribution of excited meson states to the HVP and HLbL  straightforwardly.

\begin{figure}
\includegraphics[scale=1]{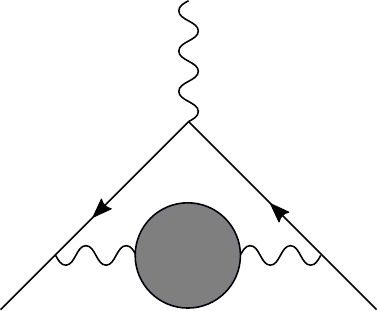}\hspace*{2em}
\includegraphics[scale=1]{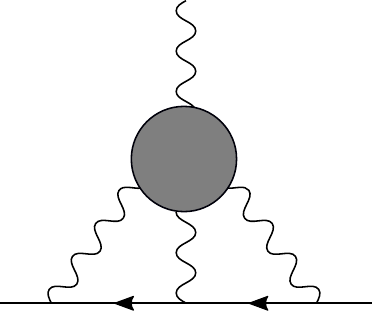}
\caption{Hadronic contributions to the muon $g-2$. The first is known as hadronic vacuum polarization (HVP), the second is hadronic light-by-light scattering (HLbL). Shaded regions denote all possible contributions of strong interactions.\label{hadronic_contribution_figure}}
\end{figure}

Another important feature of propagators in the presence of the mean gluon field is the Schwinger gauge phase factor also known as Wilson line
\begin{equation}
\mathcal{P}\,\exp\left[ig\int_x^y dz_\alpha \hat B_\alpha(z)\right],
\end{equation}
which provides explicit invariance of the effective meson action with respect to the background gauge transformations.  
Calculation of pion transition form factor $F_{\pi\gamma^*\gamma}$ within the model has demonstrated~\cite{Nedelko:2016vpj} that the Wilson lines are responsible for deviation of  $F_{\pi\gamma^*\gamma}$ from Brodsky-Lepage limit~\cite{Lepage:1980fj} at large photon virtuality. The Wilson lines were also critically important for the description of decays of vector mesons into a couple of pseudoscalar mesons simultaneously with their masses. 
It is interesting to study influence of the mean field (via confinement and Wilson lines) on the anomalous magnetic moment of muon $a_\mu$. It has to be stressed that the same values of parameters that were used for calculations of the masses and decay constants -- are employed in the present calculations.   We consider contributions to $a_\mu$ schematically shown in Fig.~\ref{hadronic_contribution_figure}, and concentrate on one-loop quark and one-meson diagrams where the role of soft mean field is seen manifestly. 

There are two main observations.  Radial excitations of vector mesons strongly enhance the contribution of hadronic vacuum polarization to  $a_\mu$. In one-meson processes the total contribution of exited states is as big as the ground-state meson contribution. The mean field strongly influences the hadronic light-by-light scattering contribution due to the Wilson line in quark propagators.  

The paper is organized as follows.
Section~\ref{section_model} contains a brief review of the model employed in this study.
Section~\ref{section_HVP} is devoted to HVP contribution to the anomalous magnetic moment of muon $a_\mu$. One-loop quark HLbL-type contribution to $a_\mu$ is considered in Section~\ref{section_HLbL}. 
The results are discussed in Section~\ref{section_discussion}.

\section{Description of the model\label{section_model}}

Motivation of the model employed in the present paper and technical details of its derivation can be found in Refs.~~\cite{Efimov:1995uz,Burdanov:1996uw,Kalloniatis:2003sa,Nedelko:2016gdk,Nedelko:2016vpj,Nedelko:2014sla,Nedelko:2020bba}. The defining property of the model is domain-structured Abelian (anti-)self-dual vacuum gluon fields which are treated nonperturbatively. The hadronization procedure of bilocal quark currents in the presence of such vacuum fields leads to the following effective nonlocal meson action (in Euclidean space):
\begin{align}
\label{meson_pf}
Z&={\cal N}
\int D\phi_{\cal Q}
\exp\left\{-\frac{\Lambda^2}{2}\frac{h^2_{\cal Q}}{g^2 C^2_\mathcal{Q}}\int d^4x 
\phi^2_{\cal Q}(x)
-\sum\limits_{k=2}^\infty\frac{1}{k}W_k[\phi]\right\},
\\
\label{effective_meson_action}
W_k[\phi]&=
\sum\limits_{{\cal Q}_1\dots{\cal Q}_k}h_{{\cal Q}_1}\dots h_{{\cal Q}_k}
\int d^4x_1\dots\int d^4x_k
\Phi_{{\cal Q}_1}(x_1)\dots \Phi_{{\cal Q}_k}(x_k)
\Gamma^{(k)}_{{\cal Q}_1\dots{\cal Q}_k}(x_1,\dots,x_k),
\\
\nonumber
\Phi_{{\cal Q}}(x)&=\int \frac{d^4p}{(2\pi)^4}e^{ipx}{\mathcal O}_{{\mathcal Q}{\mathcal Q}'}(p)\tilde\phi_{{\mathcal Q}'}(p),\quad C_\mathcal{Q}=C_J,\quad C^2_{S/P}=2C^2_{V/A}=\frac{1}{9}.
\end{align}
The condensed index $\mathcal{Q}\equiv\{aJLn\}$ denotes all quantum numbers of a meson (isospin, spin-parity, orbital and radial numbers, space-time indices for the fields with nonzero total angular momentum), and $\Lambda$ is related to the strength of the background gluon field. The orthogonal transformation ${\mathcal O}_{{\mathcal Q}{\mathcal Q}'}$ diagonalizes quadratic part of the action and thus relates physical fields $\phi_{{\mathcal Q}}$ to auxiliary fields $\Phi_{{\mathcal Q}}$ arising by virtue of the hadronization procedure. Meson masses 
are determined by the poles of the meson propagator defined from the equation
\begin{align}
\label{mass-eq}
0&=
\frac{\Lambda^2}{g^2 C^2_\mathcal{Q}}+\tilde\Gamma^{(2)}_{\cal Q}(-M^2_{\cal Q}),
\end{align}
solved with respect to the mass $M^2_{\cal Q}$ (location of the pole).
Constant $h_\mathcal{Q}$ is defined by equation
\begin{align}
1&=h^2_{\cal Q}
\left.\frac{d}{dp^2}\tilde\Gamma^{(2)}_{\cal Q}(p^2)\right|_{p^2=-M^2_{\cal Q}},
\label{hqm}
\end{align}
 solved with respect to constant $h_\mathcal{Q}$ to make the residue at the pole of meson propagator to be equal to unity. The function $\tilde\Gamma^{(2)}_{\cal Q}$ is two-point correlation function $\tilde\Gamma^{(2)}_{\cal QQ'}(p)$ diagonalized with respect to all quantum numbers:
\begin{equation}
\label{diagonalization}
\tilde\phi_{\mathcal Q}(-p)\left[\mathcal{O}^T(p)\tilde\Gamma^{(2)}(p^2)\mathcal{O}(p)\right]_{\cal QQ'}\tilde\phi_{{\mathcal Q}'}(p)=\tilde\Gamma^{(2)}_{\cal Q}(p^2)\tilde\phi_{\mathcal Q}(-p)\tilde\phi_{\mathcal Q}(p).
\end{equation}
Two-point nonlocal vertex function $\tilde\Gamma^{(2)}_{\cal QQ'}(p)$ is given by
\begin{equation}
\label{Gammak}
\Gamma^{(2)}_{{\cal Q}_1{\cal Q}_2}=
\overline{G^{(2)}_{{\cal Q}_1{\cal Q}_2}(x_1,x_2)}-
\Xi_2(x_1-x_2)\overline{G^{(1)}_{{\cal Q}_1}G^{(1)}_{{\cal Q}_2}},
\end{equation}
and analogous formulas can be written for $k$-point vertices $\Gamma^{(k)}_{\mathcal{Q}_1\dots \mathcal{Q}_k}$.
Quark loops $G^{(k)}_{{\cal Q}_1\dots{\cal Q}_k}$ are averaged over the background field with measure $d\sigma_B$:
\begin{eqnarray}
\label{barG}
&&\overline{G^{(k)}_{{\cal Q}_1\dots{\cal Q}_k}(x_1,\dots,x_k)}
=\int d \sigma_B
{\rm Tr}V_{{\cal Q}_1}\left(x_1\right)S\left(x_1,x_2\right)\dots
V_{{\cal Q}_k}\left(x_k\right)S\left(x_k,x_1\right),
\\
&&\overline{G^{(l)}_{{\cal Q}_1\dots{\cal Q}_l}(x_1,\dots,x_l)
G^{(k)}_{{\cal Q}_{l+1}\dots{\cal Q}_k}(x_{l+1},\dots,x_k)}
=
\nonumber\\
\nonumber
&&\int d \sigma_B
{\rm Tr}\left\{
V_{{\cal Q}_1}\left(x_1\right)S\left(x_1,x_2\right)\dots
V_{{\cal Q}_k}\left(x_l\right)S\left(x_l,x_1\right)
\right\}\times
\\
&&
{\rm Tr}\left\{
V_{{\cal Q}_{l+1}}\left(x_{l+1}\right)S\left(x_{l+1},x_{l+2}\right)\dots
V_{{\cal Q}_k}\left(x_k\right)S\left(x_k,x_{l+1}\right)
\right\}.
\label{barGG}
\end{eqnarray}
Here $S(x,y)$ is the quark propagator and $V_{{\cal Q}}$ are  nonlocal  meson-quark-antiquark vertices with ${\cal Q}$ - meson quantum numbers (see \cite{Nedelko:2016vpj} for details).

In order to take into account the basic properties of the mean field corresponding to almost everywhere homogeneous (anti-)self-dual Abelian gauge field represented by the statistical ensemble of domain wall networks  and still be able to perform calculations analytically, we make the following simplifications. The quark propagator and meson-quark vertices are approximated by those in the homogeneous (anti-)self-dual Abelian background field, and the statistical properties of the ensemble of domain networks are taken into account by means of averaging of the quark loops over all configurations of the  homogeneous  background fields as well as via correlators $\Xi_n$ in the background field ensemble.

Specifically, the quark loops $G^{(k)}$ are  averaged over parity-conjugated self-dual and anti-self-dual Abelian (anti-)self-dual background field and  direction (in coordinate and color spaces). Averaging over spatial directions is performed with the help of generating formula
\begin{equation}\label{averaging_over_vacuum_field}
\langle\exp(if_{\mu\nu}J_{\mu\nu})\rangle=\frac{\sin\sqrt{2\left(J_{\mu\nu}J_{\mu\nu}\pm J_{\mu\nu}\widetilde{J}_{\mu\nu}\right)}}{\sqrt{2\left(J_{\mu\nu}J_{\mu\nu}\pm J_{\mu\nu}\widetilde{J}_{\mu\nu}\right)}},
\end{equation}
where $J_{\mu\nu}$ is an arbitrary antisymmetric tensor.
Tensor $f_{\mu\nu}$ is an appropriately normalized Abelian (anti-)self-dual background field with strength $\Lambda$:
\begin{gather}
\nonumber
\hat B_\mu=-\frac{1}{2}\hat n B_{\mu\nu}x_\nu, \ \hat n = t^3\cos\xi+t^8\sin\xi,
\\
\label{b-field}
\tilde{B}_{\mu\nu}=\frac12\epsilon_{\mu\nu\alpha\beta}B_{\alpha\beta}=\pm B_{\mu\nu}, \  \hat{B}_{\rho\mu}\hat{B}_{\rho\nu}=4\upsilon^2\Lambda^4\delta_{\mu\nu},\\
\nonumber
f_{\alpha\beta}=\frac{\hat{n}}{2\upsilon\Lambda^2}B_{\alpha\beta}, \  \upsilon=\mathrm{diag}\left(\frac16,\frac16,\frac13\right), \ f_{\mu\alpha}f_{\nu\alpha}=\delta_{\mu\nu},
\end{gather}
where the upper sign in ``$\pm$'' should be taken for self-dual field, and the lower for anti-self-dual field.
Nonlocal vertices $V^{aJln}_{\mu_1\dots\mu_l}$ are given by formulas
\begin{gather}
V^{aJln}_{\mu_1\dots\mu_l}= {\cal C}_{ln}\mathcal{M}^a\Gamma^J F_{nl}\left(\frac{\stackrel{\leftrightarrow}{\cal D}^2\!\!\!
(x)}{\Lambda^2}\right)T^{(l)}_{\mu_1\dots\mu_l}\left(\frac{1}{i}\frac{\stackrel{\leftrightarrow}{\cal D}\!(x)}{\Lambda}\right),
\label{qmvert}\\
{\cal C}^2_{ln}=\frac{l+1}{2^ln!(n+l)!},\quad F_{nl}(s)=s^n\int_0^1 dt t^{n+l} \exp(st),
\nonumber\\
{\stackrel{\leftrightarrow}{\mathcal{D}}}\vphantom{D}^{ff'}_{\mu}=\xi_f\stackrel{\leftarrow}{\mathcal{D}}_{\mu}-\ \xi_{f'}\stackrel{\rightarrow}{\mathcal{D}}_{\mu}, 
\ \
\stackrel{\leftarrow}{\mathcal{D}}_{\mu}\hspace*{-0.3em}(x)=\stackrel{\leftarrow}{\partial}_\mu+\ i\hat B_\mu(x),  \ \ 
\stackrel{\rightarrow}{\mathcal{D}}_{\mu}\hspace*{-0.3em}(x)=\stackrel{\rightarrow}{\partial}_\mu-\ i\hat B_\mu(x), 
\nonumber\\
\xi_f=\frac{m_{f'}}{m_f+m_{f'}},\ \xi_{f'}=\frac{m_{f}}{m_f+m_{f'}}.
\nonumber
\end{gather}
Here $\mathcal{M}^a$ and $\Gamma^J$ are flavor and Dirac matrices corresponding to a given meson field, $\xi_f,\xi_{f'}$ provide that $x$ is the center of mass of a meson, $n,l$ are radial and orbital quantum numbers, respectively.
Radial part  $F_{nl}$ is defined by the propagator of the gluon fluctuations  charged with respect to the Abelian background, $T^{(l)}$ are irreducible tensors of four-dimensional rotation group.
Propagator of the quark with mass $m_f$ in the presence of the homogeneous Abelian (anti-)self-dual field  has the form 
\begin{align}
\label{quark_propagator}
S_f(x,y)&=\exp\left(-\frac{i}{2}\hat n x_\mu  B_{\mu\nu}y_\nu\right)H_f(x-y),
\\
\tilde H_f(p)&=\frac{1}{2\upsilon \Lambda^2} \int_0^1 ds e^{(-p^2/2\upsilon \Lambda^2)s}\left(\frac{1-s}{1+s}\right)^{m_f^2/4\upsilon \Lambda^2}
\nonumber\\
&\quad\times \left[\vphantom{\frac{s}{1-s^2}}p_\alpha\gamma_\alpha\pm is\gamma_5\gamma_\alpha f_{\alpha\beta} p_\beta
+m_f\left(P_\pm+P_\mp\frac{1+s^2}{1-s^2}-\frac{i}{2}\gamma_\alpha f_{\alpha\beta}\gamma_\beta\frac{s}{1-s^2}\right)\right],
\nonumber
\end{align}
where anti-Hermitean representation of Dirac matrices is used, and ``$\pm$'' signs are arranged in accordance with formula~\eqref{b-field}. Due to the Wilson line  translation invariance is hold only up to the large gauge transformation
\begin{align}
\label{lgtr}
S_f(x+a,y+a)=U(x)S_f(x,y)U^{+}(y), \ \  U(x)=\exp\left(-\frac{i}{2}\hat n x_\mu  B_{\mu\nu}a_\nu\right).
\end{align} 
The translation-invariant part $H_f$ of the propagator is an analytical function in the finite complex momentum plane and matches the behavior of free Dirac propagator at large Euclidean momentum.

The values of parameters of the model  given in Table~\ref{values_of_parameters} were fitted to the masses of ground-state mesons $\pi,\rho,K,K^*,J/\psi,\Upsilon,\eta'$ (see~\cite{Nedelko:2016gdk}), and the same values are employed in the present paper.
\begin{table}
\begin{tabular}{|*{6}{@{\hspace*{0.5em}}c@{\hspace*{0.5em}}|}}
\hline
$m_{u/d}$(MeV)&$m_s$(MeV)&$m_c$(MeV)&$m_b$(MeV)&$\Lambda$(MeV)&$\alpha_s$\\
\hline
$145$&$376$&$1566$&$4879$&$416$&$3.45$\\
\hline
\end{tabular}
\caption{Values of parameters introduced in Ref.~\cite{Nedelko:2016gdk} and used for calculations of mass spectrum, decay constants and form factors. The same values are employed in the present paper.\label{values_of_parameters}}
\end{table}

Electromagnetic interactions are included in this scheme with the help of expansions (see~\cite{Burdanov:1998tf,Nedelko:2016gdk})
\begin{align}
\label{propagator_electromagnetic_field}
S_f(x,y|A)&=S_f(x,y)+\sum_{n=1}^\infty \left(Q_fe\right)^n \int dz_1\cdots \int dz_n S_f(x_1,z_1)\gamma_{\mu_1}A_{\mu_1}(z_1)\cdots S_f(z_{i-1},z_i)\gamma_{\mu_i}A_{\mu_i}(z_i)\cdots S_f(z_n,y),\\
\label{vertex_electromagnetic_field}
V_\mathcal{Q}(x|A)&=V_\mathcal{Q}(x)+\sum_{n=1}^\infty e^n \int dz_1\cdots \int dz_n V_{\mathcal{Q};\mu_1\dots\mu_n}^{(n)}(x,z_1,\dots,z_n)A_{\mu_1}(z_1)\cdots A_{\mu_n}(z_n),
\end{align}
where $Q$ is a diagonal matrix of quark charges in units of electron charge $e$, and meson-photon vertices appear due to nonlocality. In the next section we give explicit examples of such interactions. Effective meson action takes the form
\begin{align}
\label{meson_pf_em}
Z&={\cal N}
\int D\phi_{\cal Q}\int DA_\mu
\exp\left\{-\frac{1}{4}\int d^4x\ F_{\mu\nu}F_{\mu\nu}-\frac{\Lambda^2}{2}\frac{h^2_{\cal Q}}{g^2 C^2_\mathcal{Q}}\int d^4x 
\phi^2_{\cal Q}(x)
-\Gamma(A)
-\sum\limits_{k=2}^\infty\frac{1}{k}W_k[\phi|A]\right\},
\\
\label{effective_meson_action_em}
W_k[\phi|A]&=
\sum\limits_{{\cal Q}_1\dots{\cal Q}_k}h_{{\cal Q}_1}\dots h_{{\cal Q}_k}
\int d^4x_1\dots\int d^4x_k
\Phi_{{\cal Q}_1}(x_1)\dots \Phi_{{\cal Q}_k}(x_k)
\Gamma^{(k)}_{{\cal Q}_1\dots{\cal Q}_k}(x_1,\dots,x_k|A),
\end{align}
where
\begin{equation*}
\Gamma(A)=\int d\sigma_B \mathrm{Tr}\log\left[1+ Qe\gamma_\mu A_\mu(x)S(x,y)\right],
\end{equation*}
and
$\Gamma^{(k)}_{{\cal Q}_1\dots{\cal Q}_k}(x_1,\dots,x_k|A)$ are obtained from $\Gamma^{(k)}_{{\cal Q}_1\dots{\cal Q}_k}(x_1,\dots,x_k)$ via substitutions
\begin{equation*}
S_f(x,y)\to S_f(x,y|A),\quad
V_\mathcal{Q}(x)\to V_\mathcal{Q}(x|A).
\end{equation*}

\section{Hadronic vacuum polarization contribution to muon $g-2$\label{section_HVP}}
We employ the following representation~\cite{Lautrup:1971jf,Blum:2002ii} for the contribution to $a_\mu$ from the first diagram in Fig.~\ref{hadronic_contribution_figure}
\begin{align}
\label{a2_HVP_representation}
a^{\text{HVP}}_\mu&=4\pi^2\left(\frac{\alpha}{\pi}\right)^2\int\limits_0^\infty dp^2
f(p^2,m_\mu^2)\tilde \Pi^{\rm R}(p^2),\\
\label{a2_HVP_representation_weight}
f(p^2,m_\mu^2)&=\frac{m_\mu^2p^2Z^3(1-p^2Z)}{1+m_\mu^2Z^2p^2},\\
\nonumber
Z&=-\left(p^2-\sqrt{p^4+4m_\mu^2p^2}\right)/2m_\mu^2p^2,
\end{align}
where $m_\mu$ is the muon mass, and $\Pi^{\rm R}(p^2)$ is related to hadronic vacuum polarization
\begin{equation}
\tilde\Pi^{\rm R}_{\mu\nu}(p)=(p^2\delta_{\mu\nu}-p_\mu p_\nu)
\tilde\Pi^{\rm R}(p^2)
\end{equation}
renormalized at $p^2=0$. HVP is defined as
\begin{equation*}
\tilde\Pi_{\mu\nu}(p)=\int d^4x e^{-ipx}\langle j_\mu(x)j_\nu(0)\rangle,
\end{equation*}
where $j_\mu=\bar\psi Q\gamma_\mu\psi$ are quark currents.

\subsection{Quark loop}
\begin{figure}
\includegraphics[scale=1]{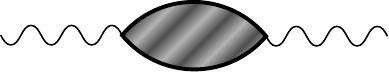}\hspace{2em}
\caption{Quark loop contributions to HVP. Solid  lines denote quark propagator~\eqref{quark_propagator}. Gradient shading indicates averaging the quark loop over configurations of the mean field.
\label{figure_HVP0}}
\end{figure}

One-loop quark contribution to HVP corresponds to the  diagram in Fig.~\ref{figure_HVP0} which is given by
\begin{equation}
\label{HVP_one-loop}
\tilde\Pi_{\mu\nu}^{(1)}(p)=\sum_f Q_f^2 \ (-1)\ {\rm reg}\int \frac{d^4l}{(2\pi)^4}\mathrm{Tr}\ 
\gamma_\mu\tilde H_f(l+p/2)\gamma_\nu\tilde H_f(l-p/2),
\end{equation}
where reg denotes some UV regularization which preserves gauge invariance. In this case with two local electromagnetic vertices, translation-noninvariant Wilson lines in quark propagators cancel each other.
Performing the loop integration and renormalization at $p^2=0$ (see Appendix~\ref{appendix_quark_loop}), we find
\begin{align*}
\tilde\Pi^{(1)\mathrm{R}}_{\mu\nu}(p)&=(p^2\delta_{\mu\nu}-p_\mu p_\nu)
\sum_f Q_f^2 \tilde\Pi_f^{\rm R}(p^2),
\\
\tilde\Pi_f^{(1)\mathrm{R}}(p^2)&=-\mathrm{Tr} \frac{1}{2\pi^2}\int_0^1ds_1\int_0^1 ds_2\frac{s_1s_2+s_1^2s_2^2/3}{(s_1+s_2)^4} \left[\frac{(1-s_1)(1-s_2)}{(1+s_1)(1+s_2)}\right]^{\frac{m_f^2}{4v\Lambda^2}}
\left[\exp\left(-\frac{s_1s_2}{s_1+s_2}\frac{p^2}{2v\Lambda^2}\right)-1\right].
\end{align*}
The above integral converges for any finite complex value of $p^2$, and its derivative is continuous, hence $\tilde\Pi_f^{(1)\mathrm{R}}(p^2)$ is an analytical function of momentum as was mentioned in the Introduction.
The expression for $\tilde\Pi_{\mu\nu}^{(1)\mathrm{R}}(p^2)$ is substituted into Eq.~\eqref{a2_HVP_representation} to obtain numbers given in Table~\ref{table_HVP_one-loop}.
\begin{table}
\begin{tabular}{|c|c|c|c|c|c||c|c|}
\hline
&$u$&$d$&$s$&$c$&$b$&total one-loop&N$\chi$QM\\
\hline
$a_\mu^\text{HVP,one-loop}$&284.7&71.2&26.4&7.2&0.2&389.6&533\\
\hline
\end{tabular}
\caption{The contributions to $a_\mu$ in units of $10^{-10}$ from one-loop quark diagram shown in Fig.~\ref{figure_HVP0}. One-loop contribution in nonlocal chiral quark model (N$\chi$QM)~\cite{Dorokhov:2004ze} is given for comparison.\label{table_HVP_one-loop}}
\end{table}

\subsection{Intermediate mesons}

\begin{figure}
\includegraphics[scale=.75]{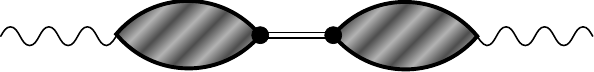}\hspace{2em}
\includegraphics[scale=.75]{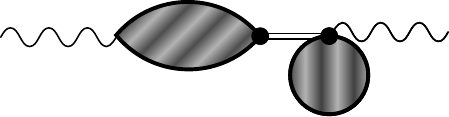}\hspace{2em}
\includegraphics[scale=.75]{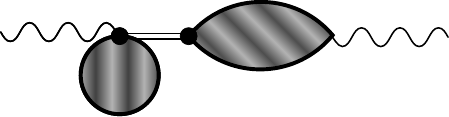}\hspace{2em}
\includegraphics[scale=.75]{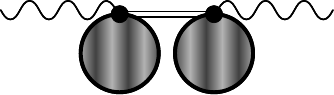}
\caption{Contributions to HVP of the first  order with respect to the number of intermediate mesons. Notation is the same as in Fig.~\ref{figure_HVP0}. 
\label{figure_HVP1}}
\end{figure}
\begin{figure}
\includegraphics[scale=1]{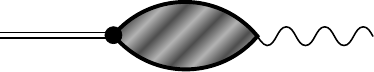}\hspace*{2em}
\includegraphics[scale=1]{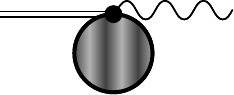}
\caption{
\label{figure_transition}
One-loop diagrams that contribute to meson-photon transition amplitude $\tilde\Gamma_{\mathcal{Q}\mu}(p)$. Notation is the same as in Fig.~\ref{figure_HVP0}. 
}
\end{figure}

Expression for the  diagrams in Fig.~\ref{figure_HVP1} is derived systematically from formula~\eqref{meson_pf} (see Appendix~\ref{appendix_HVP}) and is given by
\begin{equation}
\label{HVP_intermediate_mesons}
\tilde\Pi_{\mu\nu}^{(2)}(p)=\sum_{\mathcal{Q}}\tilde\Gamma_{\mathcal{Q};\mu}(p)\left[\frac{\Lambda^2 h_\mathcal{Q}^2}{g^2 C^2_\mathcal{Q}}+h_\mathcal{Q}^2\tilde\Gamma^{(2)}_{\cal Q}(p^2)\right]^{-1}\tilde\Gamma_{\mathcal{Q};\nu}(p),
\end{equation}
where the term
\begin{equation}
\label{meson_propagator}
\left[\frac{\Lambda^2 h_\mathcal{Q}^2}{g^2 C^2_\mathcal{Q}}+h_\mathcal{Q}^2\tilde\Gamma^{(2)}_{\cal Q}(p^2)\right]^{-1}
\end{equation}
is the propagator of a meson field (cf. Eqs.~\eqref{mass-eq} and~\eqref{hqm}). Only the meson states with quantum numbers of a photon (ground-state and  excited meson fields) contribute to Eq.~\eqref{HVP_intermediate_mesons}, while the contributions of other states are identically zero. The meson propagator includes $\Gamma_\mathcal{Q}^{(2)}$ which is given by formula~\eqref{diagonalization}, and only connected piece contributes to $\Gamma^{(2)}_{{\cal Q}_1{\cal Q}_2}$ in the case of intermediate vector mesons:
\begin{equation*}
\Gamma^{(2)}_{{\cal V}_1{\cal V}_2}=
\overline{G^{(2)}_{{\cal V}_1{\cal V}_2}(x_1,x_2)}.
\end{equation*}
In one-loop approximation, meson-photon transition amplitudes $\tilde\Gamma_{\mathcal{Q};\mu}(p)$ are given by the sum of two diagrams shown in Fig.~\ref{figure_transition} (first and second terms, correspondingly)
\begin{align*}
\tilde\Gamma_{\mathcal{Q};\mu}(p)&=h_{\mathcal{Q}}\sum_{\mathcal{Q}} O_{\mathcal{Q}'\mathcal{Q}}(p^2) \int d\sigma_B\int \frac{d^4p'}{2\pi^4}\int d^4x e^{ipx}\int d^4y e^{ip'y} \mathrm{Tr}S(y,x)V_{\mathcal{Q}'}(x)S(x,y)Q\gamma_\mu\\
&\quad+h_{\mathcal{Q}}\sum_{\mathcal{Q}'} O_{\mathcal{Q}'\mathcal{Q}}(p^2) \int d\sigma_B\int d^4z\ e^{ipz}\ \mathrm{Tr}S(x,x)V_{\mathcal{Q}';\mu}^{(1)}(x,z)\gamma_\mu.
\end{align*}

Due to nonlocality of meson-quark vertices $\tilde\Gamma_\mathcal{Q}^{(2)}$ and $\tilde\Gamma_{\mathcal{Q};\mu}(p)$ are UV-finite, and gauge invariance provides that
\begin{equation*}
p_\mu\tilde\Gamma_{\mathcal{Q};\mu}(p)=0.
\end{equation*}
Therefore, only transversal parts of meson-photon transition amplitudes and meson propagator contribute to Eq.~\eqref{HVP_intermediate_mesons}, and
\begin{equation}
\label{Pi_form-factor}
\tilde\Pi_{\mu\nu}^{(2)}(p)=(p^2\delta_{\mu\nu}-p_\mu p_\nu)\tilde\Pi^{(2)}(p^2).
\end{equation}
The explicit formulas for $\tilde\Gamma_\mathcal{Q}^{(2)}$ and $\tilde\Gamma_{\mathcal{Q}\mu}$ can be found in Ref.~\cite{Nedelko:2016gdk}, where they were used to find spectrum and transition constants of various mesons. Substituting the formulas into~\eqref{HVP_intermediate_mesons} and~\eqref{a2_HVP_representation}, we find the contributions to $a_\mu^\text{HVP}$ corresponding to various vector meson fields and their radial excitations ($n=0\dots 6$) which are given in Table~\ref{table_vector_meson_a2_l0}.
\begin{table}
\begin{tabular}{|c|c|c|c|c|c|c|c||c|c|}
\hline
&$V_0$&$V_1$&$V_2$&$V_3$&$V_4$&$V_5$&$V_6$&total&N$\chi$QM\\
\hline
$\rho_n$&$41.5$&$27$&$14.8$&$9.5$&$6$&$3.3$&$3.5$&$105.5$&\\
\cline{1-9}
$\omega_n$&$4.6$&$3$&$1.6$&$1.1$&$0.7$&$0.4$&$0.4$&$11.7$&\\
\cline{1-9}
$\phi_n$&$2.8$&$2$&$1.4$&$1$&$0.7$&$0.5$&$0.4$&$8.8$&\\
\cline{1-9}
$J/\psi_n$&$0.2$&$0.19$&$0.18$&$0.17$&$0.16$&$0.15$&$0.14$&$1.2$&\\
\cline{1-9}
total&$49.1$&$32.2$&$17.9$&$11.7$&$7.6$&$4.4$&$4.4$&$127.3$&$13$\\
\hline
\end{tabular}
\caption{The contributions to $a_\mu$ in units of $10^{-10}$ corresponding to various vector mesons and their radial excitations $V_n$. Currently, we take into account seven radial states $n$, and numerical values of contributions $V_6$ are affected by this finite number of retained radial states most notably.  Contribution of ground state vector mesons in N$\chi$QM~\cite{Dorokhov:2004ze} is shown for comparison. It can be seen that the total contribution of radial excitations is as large as the ground state itself. \label{table_vector_meson_a2_l0}}
\end{table}
Total contribution to $a_\mu^\text{HVP}$ from intermediate vector mesons is numerically smaller than the one of quark loop given in Table~\ref{table_HVP_one-loop}. Total contribution of radially excited  mesons is as large as the contribution of the ground state mesons.

\begin{table}
\begin{tabular}{|c|c|c|c|c||c|c|c|}
\hline
this paper&N$\chi$QM&ENJL&LMD&GNC&data-driven&\multicolumn{2}{|c|}{lattice QCD}\\
\hline
516.9&623(40)&750&570(170)&630(50)&693.1(4.0)&711.6(18.4)&707.5(5.5)\\
\hline
\end{tabular}
\caption{
Contributions to $a_\mu^\text{HVP}$ found in the present paper, nonlocal chiral quark model (N$\chi$QM)~\cite{Dorokhov:2004ze}, extended Nambu--Jona-Lasinio model (ENJL)~\cite{deRafael:1993za}, lowest  meson  dominance  (LMD)  approximation  to  large-N$_c$ QCD \cite{Peris:1998nj,Knecht:2003kc}, gauged nonlocal constituent quark model (GNC)~\cite{Holdom:1993ad}.
The total leading-order HVP contribution to $a_\mu$ extracted from experimental data~\cite{Aoyama:2020ynm,Davier:2017zfy,Keshavarzi:2018mgv,Colangelo:2018mtw,Hoferichter:2019mqg,Davier:2019can,Keshavarzi:2019abf} is shown in column ``data-driven''.
Average of the lattice QCD results by many authors~\cite{FermilabLattice:2017wgj,Budapest-Marseille-Wuppertal:2017okr,RBC:2018dos,Giusti:2019xct,Shintani:2019wai,FermilabLattice:2019ugu,Gerardin:2019rua,Aubin:2019usy,Giusti:2019hkz} for leading-order HVP contribution collected in Ref.~\cite{Aoyama:2020ynm} is given in column ``lattice QCD'' on the left. Central value of lattice QCD results given in Ref.~\cite{Borsanyi:2020mff} is shown on the right.
\label{table_HVP_total}}
\end{table}

The contribution to $a_\mu$ corresponding to quark loop~\eqref{HVP_one-loop} and intermediate vector mesons~\eqref{HVP_intermediate_mesons}, in the one-meson approximation, is given in Table~\ref{table_HVP_total}. The results of various other calculations are shown for the reference.  It has to be stressed that  data-driven and LQCD results take into account much more various contributions than the model calculations, the (N$\chi$QM) result accounts also for meson loop with two intermediate mesons which has been found to be three times  bigger than the one-meson part~\cite{Dorokhov:2004ze}.
It is seen that the total contribution of the quark loop and one-meson approximation underestimates the value of HVP in the present calculation.  One concludes that the contribution of two intermediate mesons has to be taken into account.
The largest of them are expected to correspond to two intermediate pions and kaons.   In the present approach, calculation of two-meson diagrams is straightforward but  somewhat bulky. Investigation of this and several
other potentially important contributions deserves a separate systematic study and  will be reported elsewhere. 

\section{Hadronic light-by-light contribution\label{section_HLbL}}
We employ the well-known projection technique and angular averages~\cite{Aldins:1970id,Knecht:2001qf,Barbieri:1974nc}
\begin{gather}
\label{a_mu_projected}
a^\text{HLbL}_\mu=-\frac{1}{48m}\mathrm{Tr}\left\{(\slashed{p}-m)[\gamma_\rho,\gamma_\sigma](\slashed{p}-m)\Gamma_{\rho\sigma}^\text{HLbL}\right\},\\
\label{combination1}
\begin{split}
\Gamma_{\rho\sigma}^\text{HLbL}(p)=\left.\frac{\partial}{\partial k_\sigma}\Gamma_\rho^\text{HLbL}(p,p')\right|_{k=0}&=e^6\int \frac{d^4q_1}{(2\pi)^4} \frac{d^4q_2}{(2\pi)^4} \frac{1}{q_1^2q_2^2(-q_1-q_2)^2}\\
&\quad\times \gamma_\mu \frac{\slashed{p}'+\slashed{q}_1-m}{(p+q_1)^2+m^2}\gamma_\lambda \frac{\slashed{p}-\slashed{q}_2-m}{(p-q_2)^2+m^2}\gamma_\nu \left.\frac{\partial}{\partial k_\rho}\Pi_{\mu\nu\lambda\sigma}(q_1,q_2,k-q_1-q_2)\right|_{k=0}
\end{split}
\end{gather}
to extract hadronic light-by-light scattering contribution to $a_\mu$. Here $\Gamma_\rho^\text{HLbL}$ corresponds to the second diagram in Fig.~\ref{hadronic_contribution_figure} (without external legs), $k$ is momentum of outgoing photon, $p,p'$ are momenta of muon in initial and final states. Hadronic light-by-light tensor is defined as
\begin{equation}
\Pi_{\mu\nu\lambda\sigma}(q_1,q_2,q_3)=-\int d^4x\ d^4y\ d^4z e^{-iq_1x-iq_2y-iq_3z}\langle 0|j_\mu(x)j_\nu(y)j_\lambda(z)j_\sigma(0)|0\rangle,
\end{equation}
where $j_\mu=\bar\psi Q\gamma_\mu\psi$ are quark currents.

\begin{figure}
\includegraphics[scale=1]{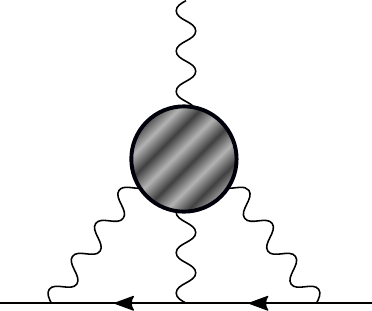}
\caption{One of the possible one-loop HLbL contributions, the other five differ by permutation of photon legs. Among these six diagrams, only three have to be calculated, the other three differ by reversed direction of the quark loop (see also~\cite{Aldins:1970id}).
\label{figure_HLbL_one-loop}}
\end{figure}

In the one-loop approximation HLbL tensor is given by the sum of six diagrams, five of which differ from the one shown in Fig.~\ref{figure_HLbL_one-loop} by permutations of photon legs:
\begin{equation}
\label{HLbL_one-loop}
\begin{split}
\Pi_{\mu\nu\lambda\sigma}(q_1,q_2,q_3)&=-\sum_f Q_f^4\int d\sigma_B \left[\int d^4x\ d^4y\ d^4z e^{-iq_1x-iq_2y-iq_3z}\right.\\
&\quad\left.\times \text{Tr}\gamma_\mu S_f(x,y) \gamma_\nu S_f(y,z) \gamma_\lambda S_f(z,0) \gamma_\sigma S_f(0,x)\vphantom{\int}\right]+\ \text{five other permutations}.
\end{split}
\end{equation}
Wilson lines in quark propagators do not cancel in this case, and momentum does not conserve in each vertex. The sum of six diagrams is finite, but there is logarithmic divergence in each diagram. HLbL tensor appears in Eq.~\eqref{combination1} in the derivative with respect to external photon momentum
\begin{equation}
\label{HLbL_derivative}
\left.\frac{\partial}{\partial k_\rho}\Pi_{\mu\nu\lambda\sigma}(q_1,q_2,k-q_1-q_2)\right|_{k=0}
\end{equation}
which makes all six terms finite~\cite{Aldins:1970id}, so it can be calculated without regularization. The external field significantly increases the size of formulas which become too cumbersome to be presented here, but the calculation is straightforward. After evaluation of traces, integrations and averaging over background field with the help of Eq.~\eqref{averaging_over_vacuum_field}, the derivative of HLbL tensor~\eqref{HLbL_derivative} is substituted into Eq.~\eqref{combination1}, which in turn is substituted into formula~\eqref{a_mu_projected}. The latter is then averaged over direction of muon momentum in Euclidean space as suggested in Ref.~\cite{Knecht:2001qf}, and analytically continued to Minkowski space $p^2\to -m_\mu^2$. The resulting integral over proper times of quark propagators~\eqref{quark_propagator}, virtualities of loop photons and the angle between photons is computed numerically. The numbers are presented in Table~\ref{HLbL_results}, column ``full''. 

\begin{table}
\begin{tabular}{|c|c|c||c|c|c|c|c|c||c|c|}
\hline
&without BF&without WL&full&N$\chi$QM&DSE&C$\chi$QM&ENJL&HLS&data-driven&QED$_\text{L}$\\
\hline
$u$&16.8&11.7&7.77&&&&&&&\\
\cline{1-4}
$d$&1&0.73&0.49&&&&&&&\\
\cline{1-4}
$s$&0.2&0.2&0.18&&&&&&&\\
\cline{1-4}
$c$&0.2&0.2&0.2&&&&&&&\\
\hline \hline
total&18.24&12.79&8.63&11(0.9)&10.7(0.2)&8.2&2.1(0.3)&0.97(1.11)&9.2(1.9)&7.87(3.06)(1.77)\\
\hline
\end{tabular}
\caption{Contributions from one-loop HLbL tensor within the model to $a_\mu$ in units of $10^{-10}$. Note that contributions of $s$ and $c$-quarks are comparable due to charge factors despite the large difference in masses.
For comparison, we present quark loop HLbL contributions in nonlocal quark chiral model (N$\chi$QM)~\cite{Dorokhov:2015psa}, Dyson-Schwinger equations (DSE)~\cite{Goecke:2012qm}, constituent chiral quark model (C$\chi$QM)~\cite{Greynat:2012ww}, extended Nambu--Jona-Lasinio model (ENJL)~\cite{Bijnens:1995xf} (result depends on cutoff, ``central value'' is shown), hidden local symmetry approach (HLS)~\cite{Hayakawa:1995ps}.
Data-driven estimate of all leading-order contributions~\cite{Aoyama:2020ynm} is given in column ``data-driven''. Result of evaluations of both QED and QCD on finite lattice~\cite{Blum:2019ugy}  is shown in column ``QED${}_\text{L}$'' (the first number in parentheses is statistical error, the second is systematic).
\label{HLbL_results}}
\end{table}

The analogous procedure is carried out for the case when Wilson lines in quark propagators are simply omitted in order to estimate their role in the final result. This is achieved by substitution (cf. Eq.~\eqref{quark_propagator})
$$S_f(x,y)\to H_f(x-y)$$
in Eq.~\eqref{HLbL_one-loop}. The numbers are given in column ``without WL'' in Table~\ref{HLbL_results}. For the column ``without BF'', background field is set to zero, and propagators $S_f(x,y)$ transform into Dirac propagators with masses given in Table~\ref{values_of_parameters}. The contributions to $a_\mu$ in this case are found with the help of formulas given in Ref.~\cite{Kuhn:2003pu}. The results show that background field and  Wilson lines affect the dynamics of light quarks most notably.

\section{Discussion\label{section_discussion}}

\begin{figure}
\includegraphics[scale=1]{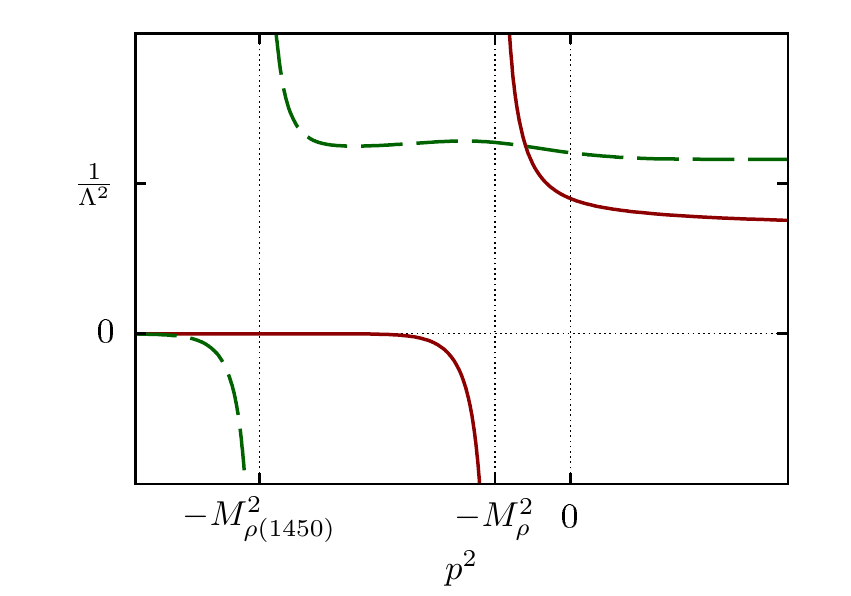}\hfill
\includegraphics[scale=1]{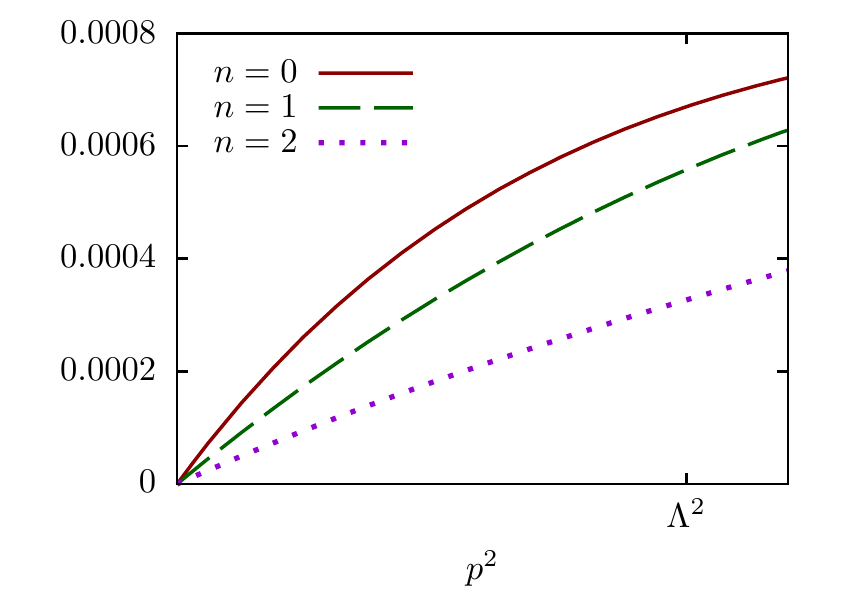}
\caption{Propagators of mesons (left panel) and corresponding contributions to $\tilde\Pi^{(2)}(p^2)$ in Eq.~\eqref{Pi_form-factor} (right panel). Propagators of ground-state $\rho$ meson and its first radial excitation $\rho(1450)$  given by formula~\eqref{meson_propagator} versus Euclidean $p^2$ are drawn with solid and dashed lines. The propagators approach $\left[ \Lambda^2 h_\mathcal{Q}^2 / g^2 C_\mathcal{Q}^2 \right]^{-1}$ at large Euclidean $p^2$ and coincide with naive propagators $\left(p^2+M_\mathcal{Q}^2\right)^{-1}$ in the vicinity of $p^2=-M_\mathcal{Q}^2$. Corresponding contributions to $\tilde\Pi^{(2)}(p^2)$ decrease with radial number $n$. \label{figure_rho_propagator}}
\end{figure}

As it has already been mentioned, rather unexpected result is that radial excitations of vector mesons turned out far from being negligible.
As it is seen from Table~\ref{table_vector_meson_a2_l0} their total contribution is even bigger than ground state meson's part.
Qualitative reason for that can be seen in highly nontrivial dependence of the meson propagator defined by the meson polarization operator, quark-meson vertices Eq.~(\ref{qmvert}) and $V\to\gamma$ transition amplitudes Eq.~(\ref{figure_transition}). Though the masses of radial excitations of vector  mesons are much larger than the mass of ground state mesons, the propagators in Euclidean domain are very different from the propagator of free vector field, and contributions of the excited states turn out to be not suppressed as much as it would be naively expected (see Fig.~\ref{figure_rho_propagator}).
It has to be stressed that all three elements---meson polarization operator, quark-meson vertices and $V\to\gamma$ amplitudes---are completely determined by the given mean field, and their overall  relevance to various meson properties have already been tested in computation of the masses, decay constant and form factors.  

\begin{figure}
\includegraphics[scale=.75]{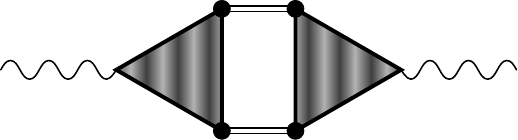}\hspace*{2em}
\includegraphics[scale=.75]{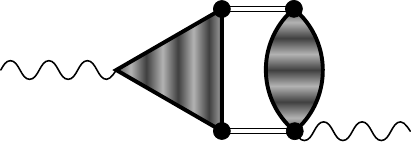}\hspace*{2em}
\includegraphics[scale=.75]{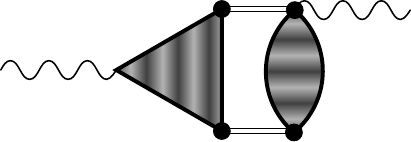} \hspace*{2em}
\includegraphics[scale=.75]{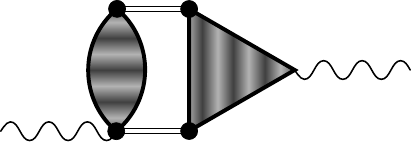}\\ \vspace*{2em}
\includegraphics[scale=.75]{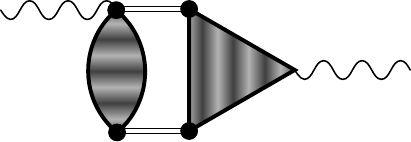}\hspace*{2em}
\includegraphics[scale=.75]{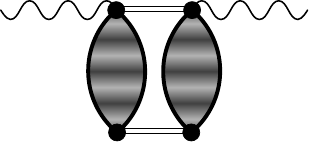}\hspace*{2em}
\includegraphics[scale=.75]{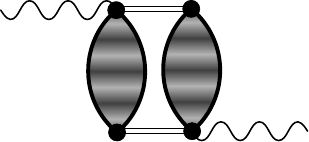} \hspace*{2em}
\includegraphics[scale=.75]{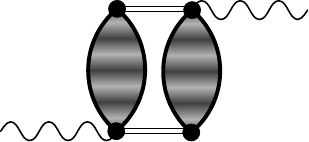}\hspace*{2em}
\includegraphics[scale=.75]{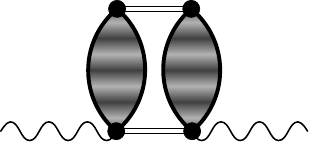}
\caption{The diagram  for the contribution of two intermediate mesons to HVP. \label{figure_HVP_2m}}
\end{figure}
Contribution coming from the lowest order in number of intermediate mesons HVP diagrams (Figs.~\ref{figure_HVP0} and~\ref{figure_HVP1}) clearly underestimates the total HVP part of $a_\mu$. However, there are many contributions  which have not been taken into account yet. First of all, these are diagrams shown in Fig.~\ref{figure_HVP_2m} with two intermediate pseudoscalar mesons including radially excited ones. There are no ad hoc arguments that their contribution must be particularly small. 
As a matter of fact, this question  addresses the role of  higher in $1/N_\mathrm{c}$ contributions.
One could also consider diagrams
that would generate imaginary part of meson propagators in Minkowski kinematics. These diagrams are
of higher order in $1/N_\mathrm{c}$  than the diagrams discussed in the article.
Diagrams with multiple internal meson lines could be also examined.
Calculation of these diagrams is quite complicated,  the result will be reported in due course.

The simplest quark loop contribution to HLbL part is quite close to both data-driven and lattice values, but only if the impact of Wilson loop is properly taken into account, see Table~\ref{HLbL_results}. 
It is clear that diagrams with intermediate   mesons, including the excited ones, have to be computed as well before coming to some definite conclusions.
\begin{figure}
\includegraphics[scale=1]{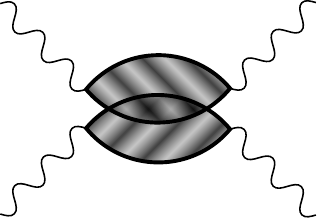}
\caption{The quark two-loop  HLbL diagram due to the mean-field correlator, quark loops are averaged jointly, in accordance with 
Eq.(\ref{barGG}).
\label{figure_HLbL_two-loop}}
\end{figure}
Besides these meson-exchange diagrams,  there exists a specific contribution illustrated by the diagram in 
Fig.~\ref{figure_HLbL_two-loop} which  corresponds to the exchange between virtual quarks 
inside loop by (infinitely) many soft gluons, represented in the domain model by the mean-field correlator. Obviously this diagram is a source of uncertainty unless the mean-field correlators are determined.    

Systematic calculations of all these and some other contributions within the domain model are straightforward but rather involved,  and their results will be reported over time.

\appendix

\section{Contributions to HVP\label{appendix_HVP}}

In order to find HVP from formula~\eqref{meson_pf} where quark fields are integrated out, it is convenient to introduce \emph{external} field $A_\mu$ and use relation
\begin{equation}
\label{HVP_functional_derivative}
\tilde\Pi_{\mu\nu}(p)=\int d^4x e^{-ipx}\langle j_\mu(x)j_\nu(0)\rangle=\int d^4x e^{-ipx} \left. \frac{1}{e^2}\frac{\delta}{\delta A_\mu(x)} \frac{\delta}{\delta A_\nu(0)} Z[A]\right|_{A=0},
\end{equation}
where $j_\mu=\bar\psi Q\gamma_\mu\psi$ are quark currents. Functional $Z[A]$ is given by (cf. Eq.~\eqref{meson_pf_em})
\begin{equation*}
Z[A]={\cal N}
\int D\phi_{\cal Q}
\exp\left\{-\frac{\Lambda^2}{2}\frac{h^2_{\cal Q}}{g^2 C^2_\mathcal{Q}}\int d^4x 
\phi^2_{\cal Q}(x)
-\Gamma(A)
-\sum\limits_{k=2}^\infty\frac{1}{k}W_k[\phi|A]\right\},\quad Z[0]=1.
\end{equation*}
Now, terms that are at most quadratic in fields $\phi_\mathcal{Q},A_\mu$ are retained in the exponential, while higher-order interactions are expanded perturbatively:
\begin{align*}
Z[A]&={\cal N}
\int D\phi_{\cal Q}\ 
\exp\left\{-\frac{\Lambda^2}{2}\frac{h^2_{\cal Q}}{g^2 C^2_\mathcal{Q}}\int d^4x 
\phi^2_{\cal Q}(x)
+\frac{1}{2}e^2\int d^4x\ d^4y\ A_\mu(x)\Pi_{\mu\nu}^{(1)}(x-y)A_\nu(y)\right.\\
&\quad\left.
-\frac{1}{2}\int d^4x\ d^4y\ \phi_{\cal Q}(x) h^2_{\cal Q}\Gamma_\mathcal{Q}^{(2)}(x-y)\phi_{\cal Q}(y)
-e\int d^4x\ d^4y\ \phi_{\cal Q}(x) h_{\cal Q}\Gamma_{\mathcal{Q};\mu}(x-y)A_\mu(y)
\right\}\\
&\quad\times\left(1+\text{higher-order interactions}\right).
\end{align*}
When higher-order terms are neglected, one performs Gaussian integration and finds
\begin{align*}
Z[A]&=
\exp\left\{\frac{1}{2}e^2\int d^4p\ \tilde A_\mu(-p)\left(\tilde\Pi_{\mu\nu}^{(1)}(p)
+\tilde\Gamma_{\mathcal{Q};\mu}(p)\left[\frac{\Lambda^2 h_\mathcal{Q}^2}{g^2 C^2_\mathcal{Q}}+h_\mathcal{Q}^2\tilde\Gamma^{(2)}_{\cal Q}(p)\right]^{-1}\tilde\Gamma_{\mathcal{Q};\nu}(p)\right)\tilde A_\nu(p)
\right\}\\
&=
\exp\left\{\frac{1}{2}e^2\int d^4p\ \tilde A_\mu(-p)\left(\tilde\Pi_{\mu\nu}^{(1)}(p)
+\tilde\Pi_{\mu\nu}^{(2)}(p)\right)\tilde A_\nu(p)
\right\}\\
&=
\exp\left\{\frac{1}{2}e^2\int d^4x\ d^4y\ A_\mu(x)\left(\Pi_{\mu\nu}^{(1)}(x-y)
+\Pi_{\mu\nu}^{(2)}(x-y)\right)A_\nu(y)
\right\}.
\end{align*}
Substituting this formula into Eq.~\eqref{HVP_functional_derivative}, one recovers diagrams shown in Figs.~\ref{figure_HVP0} and~\ref{figure_HVP1}.

\section{Quark loop contribution to HVP\label{appendix_quark_loop}}
Upon evaluating trace of Dirac matrices and performing loop momentum integration in formula~\eqref{HVP_one-loop}, one finds
\begin{align*}
(-1)\ {\rm reg}\int \frac{d^4l}{(2\pi)^4}\mathrm{Tr}\ 
\gamma_\mu\tilde H_f(l+p/2)\gamma_\nu\tilde H_f(l-p/2)=(p^2\delta_{\mu\nu}-p_\mu p_\nu)\tilde\Pi_{f1}^{(1)}(p^2)+
f_{\mu\alpha}p_\alpha f_{\nu\beta}p_\beta\tilde\Pi_{f2}^{(1)}(p^2)+\delta_{\mu\nu}\Delta.
\end{align*}
The first two structures with
\begin{align*}
\tilde\Pi_{f1}^{(1)}(p^2)&=-\mathrm{Tr} \frac{1}{(4\pi)^2}\mathrm{reg}\int_0^1ds_1\int_0^1 ds_2 \left[\frac{(1-s_1)(1-s_2)}{(1+s_1)(1+s_2)}\right]^{\frac{m_f^2}{4v\Lambda^2}}\exp\left(-\frac{s_1s_2}{s_1+s_2}\frac{p^2}{2v\Lambda^2}\right)
\frac{8s_1s_2}{(s_1+s_2)^4},\\
\tilde\Pi_{f2}^{(1)}(p^2)&=-\mathrm{Tr} \frac{1}{(4\pi)^2}\mathrm{reg}\int_0^1ds_1\int_0^1 ds_2 \left[\frac{(1-s_1)(1-s_2)}{(1+s_1)(1+s_2)}\right]^{\frac{m_f^2}{4v\Lambda^2}}\exp\left(-\frac{s_1s_2}{s_1+s_2}\frac{p^2}{2v\Lambda^2}\right)
\frac{8s_1^2s_2^2}{(s_1+s_2)^4},
\end{align*}
where the trace is over color indices, satisfy Ward identity (tensor $f_{\mu\nu}$ is antisymmetric), and the third term is equal to zero. 
In order to show this, let us represent $\Delta$ as a sum of two terms 
\begin{align*}
\Delta(p^2)&=-\Delta_1(p^2)-\Delta_2(p^2),
\\
\Delta_1(p^2)&=\mathrm{Tr} \frac{2v\Lambda^2}{(4\pi)^2}\mathrm{reg}\int\limits_0^1\int\limits_0^1\frac{ds_1 ds_2}{(s_1+s_2)^3} \left[\frac{(1-s_1)(1-s_2)}{(1+s_1)(1+s_2)}\right]^{\frac{m_f^2}{4v\Lambda^2}}\exp\left(-\frac{s_1s_2}{s_1+s_2}\frac{p^2}{2v\Lambda^2}\right)
\\
&\quad\times 
\left[ -4 \frac{p^2}{2v\Lambda^2}\frac{s_1s_2}{(s_1+s_2)}-4-4\frac{m_f^2}{2v\Lambda^2}\frac{(1-s_1s_2)(s_1+s_2)}{(1-s_1^2)(1-s_2^2)}\right],
\\
\Delta_2(p^2)&=\mathrm{Tr} \frac{2v\Lambda^2}{(4\pi)^2}\mathrm{reg} \int\limits_0^1\int\limits_0^1\frac{ds_1 ds_2}{(s_1+s_2)^3} \left[\frac{(1-s_1)(1-s_2)}{(1+s_1)(1+s_2)}\right]^{\frac{m_f^2}{4v\Lambda^2}}\exp\left(-\frac{s_1s_2}{s_1+s_2}\frac{p^2}{2v\Lambda^2}\right)
\\
&\quad\times s_1 s_2
\left[ -4 \frac{p^2}{2v\Lambda^2}\frac{s_1s_2}{(s_1+s_2)}+4-4\frac{m_f^2}{2v\Lambda^2}\frac{(1-s_1s_2)(s_1+s_2)}{(1-s_1^2)(1-s_2^2)}\right].
\end{align*}
Then $\Delta_1$ can be identically rewritten as
\begin{align*}
\Delta_1(p^2)&=4\lim_{\rho\to 1}\mathrm{Tr} \frac{2v\Lambda^2}{(4\pi)^2} \mathrm{reg}\int\limits_0^1\int\limits_0^1\frac{ds_1 ds_2}{(s_1+s_2)^3} \rho^{-1} \frac{d}{d\rho}\rho^{-1}\exp\left(-\frac{s_1s_2}{s_1+s_2}\rho \frac{p^2}{2v\Lambda^2}\right)
\\
&\quad\times\exp\left(\frac{m_f^2}{4v\Lambda^2}\ln\left[\frac{(1-\rho s_1)(1-\rho s_2)}{(1+\rho s_1)(1+\rho s_2)}\right]\right)
\\
&=4\lim_{\rho\to 1}\mathrm{Tr} \frac{2v\Lambda^2}{(4\pi)^2} \frac{d}{d\rho} \mathrm{reg}\int\limits_0^\rho\int\limits_0^\rho\frac{ds_1 ds_2}{(s_1+s_2)^3}\exp\left(-\frac{s_1s_2}{s_1+s_2}\frac{p^2}{2v\Lambda^2}+\frac{m_f^2}{4v\Lambda^2}\ln\left[\frac{(1-s_1)(1-s_2)}{(1+s_1)(1+s_2)}\right]\right)=0.
\end{align*}
The second term $\Delta_2$ is rewritten in a similar way
\begin{align*}
\Delta_2(p^2)&=4\lim_{\rho\to 1}\mathrm{Tr} \frac{2v\Lambda^2}{(4\pi)^2} \mathrm{reg}\int\limits_0^1\int\limits_0^1\frac{ds_1 ds_2 s_1s_2}{(s_1+s_2)^3} 
\rho \frac{d}{d\rho}\rho\exp\left(-\frac{s_1s_2}{s_1+s_2}\rho \frac{p^2}{2v\Lambda^2}\right)
\\
&\quad\times\exp\left(\frac{m_f^2}{4v\Lambda^2}\ln\left[\frac{(1-\rho s_1)(1-\rho s_2)}{(1+\rho s_1)(1+\rho s_2)}\right]\right)
\\
&=4\lim_{\rho\to 1}\mathrm{Tr} \frac{2v\Lambda^2}{(4\pi)^2} \frac{d}{d\rho} \mathrm{reg} \int\limits_0^\rho\int\limits_0^\rho\frac{ds_1 ds_2s_1 s_2}{(s_1+s_2)^3}\exp\left(-\frac{s_1s_2}{s_1+s_2}\frac{p^2}{2v\Lambda^2}+\frac{m_f^2}{4v\Lambda^2}\ln\left[\frac{(1-s_1)(1-s_2)}{(1+s_1)(1+s_2)}\right]\right)=0.
\end{align*}

Averaging $\tilde\Pi_{\mu\nu}(p^2)$ over background gluon field and renormalizing at $p^2=0$, one arrives at
\begin{align*}
\tilde\Pi^{(1)\mathrm{R}}_{\mu\nu}(p)&=(p^2\delta_{\mu\nu}-p_\mu p_\nu)
\sum_f Q_f^2 \tilde\Pi_f^{(1)\mathrm{R}}(p^2),
\\
\tilde\Pi_f^{(1)\mathrm{R}}(p^2)&=\tilde\Pi^{(1)\mathrm{R}}_{f1}(p^2)+\frac{1}{3}
\tilde\Pi^{(1)\mathrm{R}}_{f2}(p^2),\\
\tilde\Pi^{(1)\mathrm{R}}_{f1}(p^2)&=-\mathrm{Tr} \frac{1}{2\pi^2}\int_0^1ds_1\int_0^1 ds_2\frac{s_1s_2}{(s_1+s_2)^4} \left[\frac{(1-s_1)(1-s_2)}{(1+s_1)(1+s_2)}\right]^{\frac{m_f^2}{4v\Lambda^2}}
\left[\exp\left(-\frac{s_1s_2}{s_1+s_2}\frac{p^2}{2v\Lambda^2}\right)-1\right],\\
\tilde\Pi^{(1)\mathrm{R}}_{f2}(p^2)&=-\mathrm{Tr} \frac{1}{2\pi^2}\int_0^1ds_1\int_0^1 ds_2\frac{s_1^2s_2^2}{(s_1+s_2)^4} \left[\frac{(1-s_1)(1-s_2)}{(1+s_1)(1+s_2)}\right]^{\frac{m_f^2}{4v\Lambda^2}}
\left[\exp\left(-\frac{s_1s_2}{s_1+s_2}\frac{p^2}{2v\Lambda^2}\right)-1\right].
\end{align*}

\bibliography{references}

\end{document}